\documentstyle[prb,aps,psfig]{revtex}
\def\be{\begin{equation}}
\def\ee{\end{equation}}
\def\bc{\begin{center}}
\def\ec{\end{center}}

\begin{document}
 
\input epsf.sty
\twocolumn[\hsize\textwidth\columnwidth\hsize\csname %
@twocolumnfalse\endcsname
 
\draft
 
\widetext
 
\title{Surface critical exponents at a uniaxial Lifshitz point}
 
\author{Michel Pleimling}
 
\address{
Institut f\"ur Theoretische Physik I, Universit\"at Erlangen-N\"urnberg,
Staudtstrasse 7B3,
D -- 91058 Erlangen, Germany}

\maketitle
 
\begin{abstract}
Using Monte Carlo techniques, the surface critical behaviour of
three-dimensional semi-infinite ANNNI models 
with different surface orientations with respect to
the axis of competing interactions
is investigated. Special attention is thereby paid
to the surface criticality at the bulk uniaxial Lifshitz point encountered
in this model. 
The presented Monte
Carlo results show that the mean-field description of semi-infinite 
ANNNI models is qualitatively correct. Lifshitz point surface critical
exponents at the ordinary transition are found to depend on the surface
orientation. At the special transition point, however, no clear 
dependency of the critical exponents on the surface orientation is
revealed. The values of the surface critical exponents presented in
this study are the first estimates available beyond mean-field theory.
\end{abstract}
 
 
\phantom{.}
]
 
\narrowtext
\section{Introduction}
Critical phenomena at surfaces have been extensively studied theoretically
during the last three decades. \cite{Bin83,Die86,Die97} The surface phase
diagram of the three-dimensional semi-infinite Ising model with only nearest
neighbor couplings is well established. Introducing two different ferromagnetic 
interactions depending on whether the neighboring spins are both located
at the surface, $J_s \geq 0$, or not, $J_b > 0$, two typical scenarios are 
encountered. If the ratio of the surface coupling $J_s$ to the bulk
coupling $J_b$, $r=J_s/J_b$, is smaller than a critical value, 
$r_{sp} \approx 1.50$ for the simple cubic lattice, \cite{Bin84,Rug93} 
the system undergoes at the bulk critical 
temperature $T_{c}$ an ordinary transition, with the bulk and surface ordering
occurring at the same temperature. Beyond this critical value one first observes
the surface transition where the surface alone orders at a temperature $T_s > T_c$,
followed by the extraordinary transition
of the bulk at $T_{c}$. At the critical ratio $r_{sp}$, the special transition
point is encountered, displaying critical properties distinct from those of the
ordinary or the surface transition. The critical exponents of the different
surface universality classes are known with rather high precision.\cite{Die98}
However, far less is known on the surface critical behaviour of models
with competing interactions which is the subject of the present paper.

The axial next-nearest-neighbor Ising (ANNNI) model \cite{Ell61,Sel88} is the best
known of these models, its bulk properties being intensively investigated since
many years. Here, competition between ferromagnetic nearest neighbor and 
antiferromagnetic next-nearest neighbor couplings takes place in one direction.
The Hamiltonian of the three-dimensional version,
defined on a cubic lattice, 
may then be written in the form
\begin{eqnarray} {\cal H} &=& - J_b \sum\limits_{xyz} s_{xyz} \left( s_{(x+1)yz} + s_{x(y+1)z} + s_{xy(z+1)} \right) 
\nonumber \\ 
& &+ \kappa_b \, J_b \sum\limits_{xyz} s_{xyz}s_{xy(z+2)} \label{Gl:annni} 
\end{eqnarray}
where $J_b > 0$ and $\kappa_b > 0$ are coupling constants. The direction of competing interactions (here, the 
$z$-direction) is also called axial direction.

Spatially modulated phases due to competing interactions are observed, among others,
in magnetic systems, alloys or ferroelectrics. \cite{Sel88,Yeo88,Cum90,Sel92,Neu98}
Some compounds (as, for example, BCCD \cite{Sch98} or NaV$_2$O$_5$ \cite{Ohw01})
display very rich phase diagrams with a multitude of commensurately or
incommensurately modulated phases. A further remarkable feature of these
systems is the possible existence of a special multicritical point called
Lifshitz point. \cite{Hor75} At a Lifshitz point, a disordered, a uniformly
ordered and a periodically ordered phase become indistinguishable. 
Various systems have been shown to possess a Lifshitz point.
\cite{Sha81,Ska00,Vys92,Schr98,Ohw01}

Recently, there has been renewed interest in the uniaxial Lifshitz point
encountered in the phase diagram of the three-dimensional 
ANNNI model.\cite{Bec00,Lei00,Die00,Alb01,Shp01,Die01,Lei01,Ple01}
This strongly anisotropic equilibrium critical
point, located at $\kappa_b^L=0.270 \pm 0.004$ and $k_BT_c^L/J_b=3.7475 \pm 0.0005$, \cite{Ple01}
is characterized by the anisotropy exponent $\theta=\nu_\|^L/\nu_\perp^L $
where $\nu_\|^L$ and $\nu_\perp^L$ are the exponents of the bulk correlation
lengths parallel and perpendicular to the $z$-axis. The value $\theta \approx 0.49$
has been obtained in a second-order $\epsilon$-expansion.\cite{Die00,Shp01}
Recent thorough determinations
of Lifshitz point critical exponents using different techniques (field-theoretical
calculations \cite{Die00,Shp01}\ and Monte Carlo methods \cite{Ple01}) have yielded
excellent agreement. 
In addition, the scaling of the spin-spin correlator
at the Lifshitz point was determined \cite{Ple01} and its form shown to agree with
the prediction resulting from a generalization of conformal invariance
to the strongly anisotropic scaling at the Lifshitz point.\cite{Hen97}

Whereas the properties of the bulk ANNNI model are well established, far
less is known about the behaviour of the ANNNI model in the presence of
surfaces. Early Monte Carlo studies \cite{Sel79,Ras80} investigated ANNNI samples 
with free surfaces but did not pay special attention to
surface properties as they were exclusively interested in bulk
behaviour. Thin ANNNI films with free surfaces were studied recently and
they were shown to present a distinct phase diagram for every film thickness.\cite{Sel00}
An early attempt to study surface critical behaviour near the Lifshitz point
was undertaken by Gumbs \cite{Gum86} in the framework of mean-field theory.
A much more elaborated mean-field treatment of the semi-infinite ANNNI model
was presented recently by Binder, Frisch, and Kimball. \cite{Bin99,Fri00}
They considered two different surface orientations with respect to the axis
of competing interactions: surfaces oriented perpendicular \cite{Bin99}
or parallel \cite{Fri00} to this axis. For both cases 
the mean-field surface critical exponents at the Lifshitz points
were determined and two different
sets of critical exponents were obtained. This dependency of the
critical behaviour on the surface
orientation may be explained by the anisotropic scaling at the Lifshitz point.
These authors also predicted the existence of a surface transition at the Lifshitz
point where only the surface orders, in variance with Gumbs \cite{Gum86}
who claimed that at the Lifshitz point the surface could not order before the bulk.
It is worth noting that no results beyond mean-field theory are available
on the surface critical behaviour of the ANNNI model close to the
Lifshitz point.

In this paper I present the first Monte Carlo study of the surface critical
behaviour at the uniaxial Lifshitz point encountered in the ANNNI model.
This study uses a cluster flip
algorithm, especially designed for simulating models with competing interactions,
which has been successfully employed in a recent investigation of
the bulk ANNNI model.\cite{Ple01}

The paper is organized in the following way. The next Section is devoted to the
presentation of semi-infinite ANNNI models with two different surface orientations.
Some details of the numerical method are also presented. Section III deals with
surfaces perpendicular to the axis of competing interactions, paying special
attention to the surface phase diagram and to the determination of surface
critical exponents. Surfaces oriented parallel to this special direction
are treated in Section IV. A short summary and outlook conclude the paper.

\section{Models and method}
Due to the anisotropy of the ANNNI model, surfaces with different orientations
are not equivalent. The two surface orientations with respect to the axial direction
considered in this paper are the followings (see Figure 1): surfaces perpendicular
to the axis of competing interactions (case A) and surfaces parallel to this
axis (case B). The same orientations were treated in recent mean-field
studies.\cite{Bin99,Fri00}

For case A modified surface couplings with strength $J_s$ connecting neighboring surface spins
are introduced in addition to the usual ANNNI interactions, see Eq.\ (1) and
Figure 1a. Three different
scenarios have to be distinguished, depending on the value of the bulk competition
parameter $\kappa_b$. When $\kappa_b$ is smaller than the Lifshitz point
value $\kappa_b^L$, the bulk undergoes
a second order phase
transition between the disordered high temperature phase and the ordered,
ferromagnetic, low temperature phase at the critical temperature $T_c(\kappa_b)$. 
This phase transition belongs to the 
universality class of the 3D Ising model. Consequently, the surface phase diagram will
resemble that of the 3D semi-infinite Ising model, with a possible shift of the location of
the special transition point, $r_{sp}(\kappa_b)$, as function of $\kappa_b$.\cite{Bin99} 
At the Lifshitz point, $\kappa_b=\kappa_b^L$,
the recent mean-field treatment yields
a surface phase diagram for the semi-infinite ANNNI model
similar to the Ising model, but with critical
exponents which differ from those of the Ising model. The conjectured existence \cite{Bin99}
of a surface transition at the Lifshitz point, 
and therefore of a special transition point, is in agreement with
the Monte Carlo results presented in the next Section. Finally, for strong 
axial next-nearest neighbor bulk couplings, $\kappa_b > \kappa_b^L$, the bulk
phase transition belongs to the universality class of the 3D $XY$ model.
Therefore, for weak surface couplings, the ordinary transition critical behaviour
should be identical to that of the three-dimensional semi-infinite 
$XY$ model,\cite{Lan89} whereas for strong surface couplings the surface effectively
decouples from the bulk and a two-dimensional bulk Ising critical behaviour
is expected at the surface transition. These two critical lines meet
at a multicritical point which should have interesting properties.
However, I will in this study not consider the latter case as I am mainly interested
in the surface critical behaviour at the Lifshitz point.

Case B with surfaces oriented parallel to the axial direction (see Figure 1b) is
the most interesting but also the most demanding case. Introducing modified
nearest neighbor, $J_s$, {\it and} axial next-nearest neighbor couplings,
$\kappa_s \geq 0$, in the surface layer leads to intriguing and very
complex situations. For example, a multicritical point shows up where 
an ordinary transition with a modulated bulk 
meets a surface transition to a
floating incommensurate phase in the surface layer. 
In this study I will not discuss
these pecularities but concentrate exclusively on the case
$\kappa_s \leq \kappa_b^L$. For these
values of the surface competition parameter $\kappa_s$
the corresponding two-dimensional ANNNI model presents a transition, belonging
to the 2D Ising universality class, from the disordered phase to the low temperature
ferromagnetic phase. \cite{Bea85}

Surface quantities are determined by simulating ANNNI films consisting of
$L_x \times L_y \times L_z$ spins, with free boundary conditions perpendicular to
the surfaces and periodic boundary conditions otherwise. Hence, for case A 
the system consists of $L_z$ layers with $L_x \times L_y$ spins per layer, whereas
for case B the samples are formed by $L_y$ layers containing $L_x \times L_z$ spins.
The semi-infinite models are then obtained in the limit $L_x$, $L_y$,
$L_z \longrightarrow \infty$. Some care is needed when choosing the shape
of the finite ANNNI films in the vicinity of the Lifshitz point. On the one hand,
the special finite-size effects coming from the anisotropic scaling at this
point \cite{Bin89} are best taken into account \cite{Ple01} by choosing 
anisotropic samples with an increased number of sites perpendicular to 
the axial direction. On the other hand, computation of surface quantities 
is usually done by simulating samples where the linear dimension of the surfaces
exceeds the film thickness.\cite{Ple98} This does not pose any problem for case
A, as here the axial direction coincides with the direction perpendicular to the 
surfaces. Therefore, for this case, systems of anisotropic shape with $L_z$ layers
and $L_x^2$ spins per layer are simulated, $L_z$ ranging from 10 to 120 and $L_x$
from 10 to 60. For case B, however, surfaces are parallel to the axial direction.
In order to balance the competing finite-size effects, I have chosen to
simulate cubes with $L_x^3$ spins, $L_x$ ranging from 10 to 80. For comparison,
a few simulations have also been done for samples with anisotropic shapes.

Critical phenomena are best studied numerically with non-local Monte Carlo methods.
Recently, we proposed a cluster-flip algorithm, based on the one-cluster flip
algorithm of Wolff, \cite{Wol89} especially designed for simulating spin systems
with competing interactions. \cite{Ple01,Hen01} Below, this algorithm, which combines
the cluster algorithm approaches for systems with long-range ferromagnetic couplings \cite{Lui95}
with that for systems with nearest neighbor random couplings,\cite{Dot91} is
reformulated in order to take modified surface couplings into account.

Let $i$ be a lattice site characterized by the spin $s_i$ and already belonging
to the cluster we iteratively build up. A nearest neighbor site $j$ with spin
$s_j$ is added to the cluster with probability 
\begin{equation}
\frac{1}{2} \left(1 + \mbox{sign} \left(
s_i s_j \right) \right) \left( 1 - \exp \left[ - 2 J_s /(k_B T) \right] \right)
\end{equation}
if sites $i$ and $j$ are both located in the surface layer, and with 
probability
\begin{equation}
\frac{1}{2} \left(1 + \mbox{sign} \left(
s_i s_j \right) \right) \left( 1 - \exp \left[ - 2 J_b /(k_B T) \right] \right)
\end{equation}
otherwise. A lattice site $k$ with spin $s_k$ axial next-nearest neighbor
to $i$ is included with probability
\begin{equation}
\frac{1}{2} \left( 1 - \mbox{sign} \left(
s_i s_k \right) \right) \left( 1 - \exp \left[ -  2 J_s \kappa_s 
/(k_B T) \right] \right)
\end{equation}
for case B if both spins are surface spins, and with probability
\begin{equation}
\frac{1}{2} \left( 1 - \mbox{sign} \left(
s_i s_k \right) \right) \left( 1 - \exp \left[ -  2 J_b \kappa_b
/(k_B T) \right] \right)
\end{equation}
otherwise. The final cluster, which is flipped as a whole, presents two
pecularities which are worth mentioning.\cite{Hen01} First, it contains spins of
both signs, in variance with the Wolff cluster method, where all spins
belonging to a cluster to be flipped have the same sign. Second, the cluster spins are
not always connected by nearest neighbor bonds, which is again different from
the traditional cluster flip method. This algorithm works
well in the vicinity of the Lifshitz point, as demonstrated in the recent
numerical study of the bulk Lifshitz point critical behaviour.\cite{Ple01}

During the simulation, layer-dependent quantities are computed. 
I discuss these quantities here only for case A, the corresponding quantities for case B
are obtained by replacing $z$ by $y$ and $L_z$ by $L_y$.
Of great interest when studying surface properties are
the magnetization per layer 
\begin{equation}
m(z)= \frac{1}{L_x L_y} \left< \left| \sum\limits_{xy} \, s_{xyz} \right| \right>
\end{equation}
and
the susceptibility per layer 
\begin{equation}
\chi(z)=\frac{L_x L_y}{k_BT} \left[ \left< \left( \frac{1}{L_x L_y} \sum\limits_{xy} \, s_{xyz} 
\right)^2 \right> - \left( m(z) \right)^2 \right] .
\end{equation}
The surface magnetization is then
$m_1=m(z=1)=m(z=L_z)$, whereas the response of the surface magnetization to 
a surface field is $\chi_{11}=\chi(z=1)=\chi(z=L_z)$. From the profiles $m(z)$
and $\chi(z)$ one also obtains the surface excess quantities
\begin{equation}
m_s=\sum\limits_{z=1} \left(m_b- m(z) \right)
\end{equation}
and
\begin{equation}  
\chi_s=\sum\limits_{z=1} \left( \chi_b - \chi(z) \right)
\end{equation}  
where $m_b$ and $\chi_b$ are the bulk magnetization and the bulk susceptibility.
The response of the surface to a bulk field, $\chi_1$, the energy, the
specific heat, and the Binder cumulant have also been computed.

Thermal averages are extracted by generating, after equilibration,
$5 \times 10^5$ clusters, and error bars are obtained by averaging over at
least ten different realizations using different random numbers.

\section{Surfaces perpendicular to the axial direction}
Semi-infinite ANNNI models with surfaces oriented perpendicular to the 
direction of competing interactions exhibit
a crossover from Ising surface critical behaviour for $\kappa_b=0$ to
Lifshitz point surface criticality for $\kappa_b=\kappa_b^L$. For the
purpose of studying this crossover, simulations were not only done
for the value of $\kappa_b$ at the Lifshitz point, $\kappa_b^L=0.27$, but also at 
$\kappa_b=0.15$ and $\kappa_b=0.24$. 
The corresponding critical temperatures are listed in Table I.
Note that the same values of 
$\kappa_b$ have been used in a study of ANNNI bulk critical behaviour.\cite{Kas85}

Various values of the surface coupling strength, $J_s$, have been considered,
with $J_s$ ranging from 0 to $3 J_b$. The resulting surface phase diagrams are displayed
in Figure 2, together with the corresponding phase diagram of the semi-infinte
Ising model, $\kappa_b=0$. The horizontal lines indicate the bulk critical temperatures
which decrease with increasing $\kappa_b$. One also observes a shift of the location
of the special transition point to lower values of $r=J_s/J_b$ for larger values
of the competition parameter $\kappa_b$, in accordance with the recent mean-field results.\cite{Bin99}
This shift is mainly due to the decrease of the bulk critical temperature.
The different 
surface transition lines in Figure 2 merge for $J_s \gg J_b$ and form
one line independent of
the value of $\kappa_b$. This follows from the fact that for very
strong surface couplings the surface effectively decouples from the bulk. 
At the Lifshitz point value
$\kappa_b=0.27$, the phase diagram is also similar to the Ising case, with an ordinary transition,
a surface transition, an extraordinary transition, and a special transition point.
This is in agreement with the mean-field treatment of Ref.\ \cite{Bin99}, but disagrees 
with the results obtained in Ref.\ \cite{Gum86}. Based on our data, the 
special transition point of the semi-infinite ANNNI model 
at the Lifshitz point is located at $r_{sp}^L = 1.15 \pm 0.05$.

Crossover phenomena showing up in the surface magnetization may be analysed by
plotting the effective exponent
\begin{equation}
\beta_{1,eff}= \mbox{d} \ln m_1 / \mbox{d} \ln t
\end{equation}
where $t=\left( T_c - T \right)/T_c$ is the reduced temperature. On approaching $T_c$,
$t \longrightarrow 0$, $\beta_{1,eff}$ becomes the critical exponent $\beta_1$ of the
surface magnetization. In the following, data unaffected by finite-size
effects are usually displayed, finite-size dependences 
being circumvented by adjusting the size
of the sample.\cite{Ple98}

Effective surface magnetization exponents obtained for $\kappa_b=0$, 0.15, and 0.24, with $J_s=J_b$
are shown in Figure 3. Clearly, for a fixed value of $t$, $\beta_{1,eff}$ decreases with increasing
$\kappa_b$. In the limit $t \longrightarrow 0$, however, the effective exponents all tend asymptotically to the 
critical surface magnetization exponent $\beta_1 \approx 0.80$ of the semi-infinite Ising model 
at the ordinary transition,\cite{Ple98} in accordance with the Ising character of the ANNNI bulk transition for
$\kappa_b < \kappa_b^L$.\cite{Kas85} The observed increase of the corrections to scaling with $\kappa_b$
reflects not only the closer proximity of the bulk Lifshitz point but also the reduced distance
of $r = J_s/J_b = 1$ to the special transition point $r_{sp}(\kappa_b)$. This is illustrated in
Figure 3 by the data obtained for $\kappa_b=0.24$ with $J_s=0.5 J_b$ where corrections to scaling
are greatly reduced compared to the case $J_s=J_b$

Some of the results obtained for $\kappa_b = \kappa_b^L$, with $r < r_{sp}^L$, are displayed
in Figures 4 and 5. For this choice of the interactions, the surface undergoes an ordinary transition
at the Lifshitz point critical temperature $T_c^L$. Both the magnetization per layer, shown in 
Figure 4, and the susceptibility per layer (not shown) vary non-monotonously close to $T_c^L$.
Whereas a non-monotonic behaviour of the layer susceptibility is also observed close to the
ordinary transition in thick Ising films, the layer magnetization increases
monotonically in Ising films from its surface value to the bulk value.\cite{Ple98} The non-monotonic behaviour
of the layer magnetization close to the Lifshitz point ordinary transition may be
explained in the following way. 
Inside the bulk the value of the 
magnetization mainly results from the balancing
of competing influences in axial direction. Close to the surface, however, some negative contributions
are missing, which then yields a tendency to stronger local ordering, resulting in a maximum 
of the layer magnetization near the surface.
Similar non-monotonic profiles also show up in the semi-infinite critical Ising model
in the presence of a weak surface field.\cite{Rit96}
It should be noted that in our case the appearance of non-monotonic
profiles depends strongly on the value of the competition parameter $\kappa_b$. This behaviour is
not observed for the studied values of $\kappa_b$ smaller than $\kappa_b^L$.
For $\kappa_b=\kappa_b^L$ the maximum in the magnetization profile shows up
for all values of the surface coupling $J_s$ leading
to the ordinary transition. When the temperature is increased, the maxima both of the layer 
magnetization and of the layer susceptibilty are shifted towards the center of the system.

The surface magnetizations and the corresponding effective exponents obtained for two different
values of the coupling ratio $J_s/J_b$ are compared in Figure 5. As expected, the amplitudes
(see Figure 5a) and the corrections to scaling (see Figure 5b) differ. It is only close to $T_c$,
for $t \leq 0.07$, that the effective exponents (and therefore the corrections to scaling)
become identical. Furthermore, the effective exponents then vary almost linearly with temperature.  
A linear extrapolation yields the value $\beta_1^L = 0.618 \pm 0.005$ for the surface magnetization critical exponent
at the Lifshitz point ordinary transition. This value is clearly smaller than the value $\beta_1 = 0.80 \pm 0.01$
obtained at the ordinary transition of the semi-infinite Ising model.\cite{Ple98} It is worth noting that
mean-field approximation yields the common value $\beta_1^{MF}=1$ for both cases.

Estimates of Lifshitz point surface critical exponents at the ordinary transition
are gathered in Table II. Hereby, the critical exponent $\gamma_1^L = 0.84 \pm 0.05$ of the susceptibilty $\chi_1$,
i.e.\ of the response of the surface to a bulk field, can be obtained by analysing the
corresponding effective exponent in the same manner as discussed previously. The response
of the surface to a surface field, $\chi_{11}$, however, is finite at $T_c^L$, with a cusp-like
singularity when approaching the critical point from high temperatures.
Instead of trying to extract the corresponding critical exponent $\gamma_{11}^L$ directly
from $\chi_{11}$, it is easier to study the derivative $\chi_{11}'=\mbox{d} \chi_{11}/\mbox{d}T$
which in our case diverges on approach to the critical point
with the power law $\chi_{11}' \sim
(-t)^{-\gamma_{11}^L -1}$. This then yields the value $-0.06 \pm 0.02$
for $\gamma_{11}^L$.

Further Lifshitz point surface critical exponents are obtained by analysing the power law
behaviour of the excess quantities $m_s$ and $\chi_s$, see Eq.\ (8) and (9), close to the critical point:
$m_s \sim t^{\beta_s^L}$ and $\chi_s \sim t^{- \gamma_s^L}$. Based on our data, the excess quantities
critical exponents are estimated in the present case to be $\beta_s^L = -0.14 \pm 0.04$ and 
$\gamma_s^L = 1.69 \pm 0.07$.

A closer inspection of Table II reveals that various scaling relations are fulfilled, thus
demonstrating the reliability of our estimates. Indeed, inserting the Lifshitz point bulk
critical exponents $\gamma_b^L= 1.36 \pm 0.03$ and $\beta_b^L = 0.238 \pm 0.005$ \cite{Ple01} as well as 
$\nu_\|^L = 0.348$,\cite{Shp01}
the following scaling relations are readily verified:
\begin{equation} \label{gl:skal1}
\beta_s^L=\beta_b^L - \nu_\|^L
\end{equation}
\begin{equation} \label{gl:skal2}
\gamma_s^L=\gamma_b^L+\nu_\|^L
\end{equation}
\begin{equation} \label{gl:skal3}
2 \gamma_1^L - \gamma_{11}^L = \gamma_s^L
\end{equation}

Lifshitz point surface quantities have also been studied near the special transition
point with $J_s=1.15 J_b$, see above. At temperatures below $T_c^L$ magnetization 
profiles are extremely flat: only very close to the surface does one observe
a slow increase of the layer magnetization which reaches a maximum in the layer 
immediately below the
surface layer. The surface magnetization itself is slightly smaller than the bulk magnetization.
Estimated values of Lifshitz point surface critical exponents at the special transition point
are listed in Table II. 
As both the bulk and the surface become critical at the special transition point,
the responses of the surface to a bulk field as well as to a surface field diverge.
The corresponding critical exponents are estimated to be $\gamma_{1,sp}^L=1.28 \pm 0.08$
and $\gamma_{11,sp}^L=0.76 \pm 0.05$ where effective exponents have again been analysed.
For the surface magnetization critical exponent the value $\beta_{1,sp} = 0.22 \pm 0.02$
is obtained.
Error bars given for the special transition point critical
exponents include the uncertainty in the localisation of this multicritical point.

It is worth noting that the Lifshitz point surface critical exponent at the special
transition point given in Table II agree within the error bars with the estimates \cite{Rug95,Die98}
of the corresponding exponents for the semi-infinite Ising model. The change from 
Ising to Lifshitz point bulk critical behaviour when $\kappa_b \longrightarrow \kappa_b^L$
seems to have only a minor impact on the computed surface critical behaviour at the special
transition point. However, this may not hold for the crossover exponent, which
describes the merging of the surface transition line with this point,\cite{Bin83}
as different values are obtained in mean-field approximation 
for the Ising and for the Lifshitz point case.\cite{Bin99}
In the present work, no attempt was made to determine the crossover exponent.

\section{Surfaces parallel to the axial direction}
Two different approaches have been chosen for studying the critical behaviour of
surfaces oriented parallel to the axial direction. In the first approach the value of
the surface competition parameter $\kappa_s$ is set to zero so that only
the strength of the surface nearest neighbor coupling $J_s$ is varied. In the second
approach the bulk and surface competition parameters have the same value, $\kappa_s =
\kappa_b$, with $J_s$ ranging from 0 to $6 J_b$. 
Simulations done for the latter case may be viewed as a preparatory
work for the study of the more complex multicritical points mentioned in Section II.
In this work I mainly discuss Lifshitz point surface criticality, i.e.\
$\kappa_b = \kappa_b^L$, at the ordinary transition and at the special transition
point. 

First one has to note that the critical value of the coupling ratio 
$r_{sp}=J_s/J_b$, needed for the surface to get critical by itself, now
depends both on the bulk and on the surface competition parameters.
For example, at the bulk Lifshitz point $r^L_{sp}=1.75 \pm 0.05$ for
$\kappa_s=0$, whereas for $\kappa_s=\kappa_b^L$ the special transition point
is tentatively located at $r^L_{sp} \approx 4.3$. This remarkable shift of the critical
coupling ratio to larger values in the latter case may be explained by a 
stronger weakening of the effectice surface couplings, due to the competing
surface interactions, than of the effective bulk couplings.

At the bulk Lifshitz point, the magnetization per layer increases monotonically
from its surface value to its bulk value, as shown in Figure 6. This is in
marked contrast to the behaviour encountered when the surface is perpendicular
to the axial direction, as discussed in the previous Section, see Figure 4.
For the surface orientation considered here, the layers are parallel to the direction of
competing interactions, so that no axial contributions are missing close to 
the surface and no tendency to stronger local ordering is expected.

Lifshitz point surface critical exponents at the ordinary transition, which 
have been computed in a similar way as for case A, are gathered in Table III.
These values should be compared with the corresponding values 
of case A surfaces listed in Table II.
Figure 7 shows effective exponents derived from the surface magnetizations
obtained by simulating samples with different surface orientations.
Clearly, both types of surfaces yield different asymptotic values $\beta_1^L$ for
the effective exponent, thus demonstrating that the values of surface
critical exponents depend on the surface orientation.
Different values are also obtained for the critical exponents $\gamma_{11}^L$,
$\beta_s^L$, and $\gamma_s^L$. For $\gamma_1^L$ the situation is not so clear
as rather similar values follow from the analysis of effective exponents.
One may note that mean-field theory yields for $\gamma_1^L$ the same value
for both surface orientations.\cite{Bin99,Fri00}

Whereas the scaling relation (\ref{gl:skal3}) is fulfilled independently
of the orientation of the surface, the scaling relations (\ref{gl:skal1})
and (\ref{gl:skal2}) have to be modified for surfaces parallel to the
axial direction. Indeed, the behaviour of excess quantities is governed
close to a bulk critical point by the bulk correlation length along
the direction perpendicular to the surface. The Lifshitz point being
an anisotropic critical point characterized by two 
correlation lengths diverging with different critical exponents,
$\nu_\perp^L$ respectively $\nu_\|^L$ has to be used for
case B respectively A surfaces.
This then yields for surfaces parallel to the axial direction the scaling relations
\begin{equation} \label{gl:skal4}
\beta_s^L=\beta_b^L - \nu_\perp^L
\end{equation}
and
\begin{equation}\label{gl:skal5}
\gamma_s^L=\gamma_b^L+\nu_\perp^L.
\end{equation}
Using $\nu_\perp^L = 0.709$,\cite{Shp01} a nice agreement with the numerical
estimates for $\beta_s^L$ and $\gamma_s^L$ is observed.

I also studied the Lifshitz point surface critical behaviour of surfaces oriented parallel to the
axial direction close to the special transition point, choosing $\kappa_s=0$ and $J_s=1.75 J_b$,
with $\kappa_b=\kappa_b^L$. The estimates for the critical exponents $\beta_{1,sp}^L$,
$\gamma_{1,sp}^L$, and $\gamma_{11,sp}^L$ have been included in Table III. Again, the quoted
error bars take into account
the sample averaging as well as the uncertainty in the location of the
special transition point. Interestingly, the values of the critical exponents are
very similar to the values obtained for the other type of surfaces considered in Section III
and quoted in Table II. This is a strong indication that at the bulk Lifshitz point
the surface critical behaviour at the special transition point may not depend
on the surface orientation.

\section{Conclusion and outlook}
Whereas surface criticality of models with only ferromagnetic couplings has been
the subject of numerous investigations, studies of surface critical behaviour in 
systems with competing interactions are very scarce. The aim of the present work
is to fill some gaps by presenting the first Monte Carlo study of surface critical
behaviour in semi-infinite three-dimensional ANNNI models, paying special
attention to the ordinary transition as well as to the special transition point,
both at the bulk Lifshitz point. Prior works \cite{Gum86,Bin99,Fri00} exclusively
focused on the mean-field treatment of this problem.

Overall, the numerical results, which have been obtained by studying different
surface orientations with respect to the axis of competing interactions, are in
qualitative agreement with results obtained in the recent mean-field treatments.\cite{Bin99,Fri00}
The location of the special transition point is a function of the bulk competition
parameter as well as of the surface competition parameter, when present. At the bulk
Lifshitz point, the surface diagram is similar to that of the semi-infinite $3D$ Ising model,
with a surface transition line ending in a special transition point. This is in accordance
with recent mean-field results,\cite{Bin99,Fri00} but disagrees with the early treatment
of Gumbs.\cite{Gum86} Finally, at the bulk Lifshitz point, surface critical exponents at 
the ordinary transition depend on the orientation of the surface, as predicted by mean-field theory.

The main results of our study are the Monte Carlo estimates of surface critical exponents
at the bulk Lifshitz point gathered in Tables II and III. 
This is the first time that values for these exponents have been determined
numerically.
Whereas at the ordinary transition surface critical exponents depend
on the surface orientation, the computed values at the special transition point 
do not show a clear dependency on the orientation of the surface with respect to the 
axial direction. Furthermore, these values agree rather well with the values obtained in 
the Ising model at the special transition point. In order to clarify the situation
at this multicritical point renormalization group calculations are desirable. However,
with regard to the immense technical difficulties already encountered in the
field-theoretical computation of bulk critical exponents at the Lifshitz point,
\cite{Die00,Alb01,Shp01,Die01,Lei01} this seems to be a formidable task not easily achievable
in the near future.

A host of open and interesting problems remains for future work. In the present study,
surface criticality has only been investigated for
values of the bulk competition parameter $\kappa_b \leq \kappa_b^L$,
thus neglecting the interesting case when the bulk undergoes a phase transition to 
a modulated structure. Similarly, for surfaces parallel to the axial direction,
the surface competition parameter $\kappa_s$ has been chosen in such a way that at the surface
transition a uniformly ordered phase appears in the surface layer. However,
for large values of $\kappa_s$ one will observe a surface transition to a floating
incommensurate phase in the surface layer. Various interesting scenarios follow,
one of the most intersting being a multicritical point where a surface transition line with a floating
phase in the surface layer meets an ordinary transition line with a modulated bulk.

A further interesting question concerns the scaling form at the Lifshitz point of
spin-spin correlation functions along and
perpendicular to the surface. A recent generalization of conformal invariance to the 
strong anisotropic scaling at the Lifshitz point\cite{Hen97} has yielded predictions for the 
scaling form of bulk two-point correlators which were found to agree perfectly
with Monte Carlo results.\cite{Ple01} Similarly, predictions for 
two-point correlation functions in semi-infinite systems
at the Lifshitz point
can be obtained in the framework of
this theory and may then be compared to numerical data. Work along the sketched lines
is planned for the future.

\acknowledgments
I thank Malte Henkel for useful discussions, and him and Alfred H\"{u}ller
for a critical reading of the manuscript.


\begin{table} \label{tabl1}
\caption{Bulk critical temperatures obtained for different values of
the bulk competition parameter $\kappa_b$.}
\begin{tabular}{|l||l|l|l|}
$\kappa_b$ & 0.15 & 0.24 & 0.27 \\
\hline
$k_BT_c/J_b$ & $4.131 \pm 0.004$ & $3.865 \pm 0.003$ & $3.7475 \pm 0.0005$ \\
\end{tabular}
\end{table}

\begin{table} \label{tabl2}
\caption{Surface critical exponents at the bulk Lifshitz point obtained
for the $3D$ semi-infinite ANNNI model with the surface layer oriented
perpendicular to the axial direction. OT-MF: mean-field exponents
at the ordinary transition \cite{Bin99}, OT-MC: Monte Carlo values
at the ordinary transition, SP-MC: Monte Carlo values at the
special transition point.
The numbers in brackets give the estimated error in the last digit.}
\begin{tabular}{|l||l|l|l|l|l|}
 & $\beta_1^L$ & $\gamma_1^L$ & $\gamma_{11}^L$ & $\beta_s^L$ & $\gamma_s^L$ \\
\hline
OT-MF & 1 & 1/2 & -1/4 & 1/4 & 5/4 \\
\hline
OT-MC & $0.618(5)$ & 0.84(5) & -0.06(2) & -0.14(4) & 1.69(7) \\
\hline\hline
& $\beta_{1,sp}^L$ & $\gamma_{1,sp}^L$ & $\gamma_{11,sp}^L$ & &  \\
\hline
SP-MC & 0.22(2) & 1.28(8) & 0.76(5) & & \\
\end{tabular}
\end{table}

\begin{table} \label{tabl3}
\caption{The same as in Table II, but now for the $3D$ semi-infinite ANNNI model with
the surface layer oriented parallel to the axial direction. Mean-field values
are extracted from Ref.\ \cite{Fri00}. The mean-field values 
of the critical exponents $\beta_s^L$ and $\gamma_s^L$, which are not given 
in Ref.\ \cite{Fri00}, follow from the
assumption that the scaling relations (\ref{gl:skal4}) and (\ref{gl:skal5}) hold.}
\begin{tabular}{|l||l|l|l|l|l|}
 & $\beta_1^L$ & $\gamma_1^L$ & $\gamma_{11}^L$ & $\beta_s^L$ & $\gamma_s^L$ \\
\hline
OT-MF & 1 & 1/2 & -1/2 & 0 & 3/2 \\
\hline
OT-MC & $0.687(5)$ & 0.82(4) & -0.29(6) & -0.46(3) & 1.98(8) \\
\hline\hline
& $\beta_{1,sp}^L$ & $\gamma_{1,sp}^L$ & $\gamma_{11,sp}^L$ & &  \\
\hline
SP-MC & 0.23(1) & 1.30(6) & 0.72(4) & & \\
\end{tabular}
\end{table}

\begin{figure}
\centerline{\psfig{figure=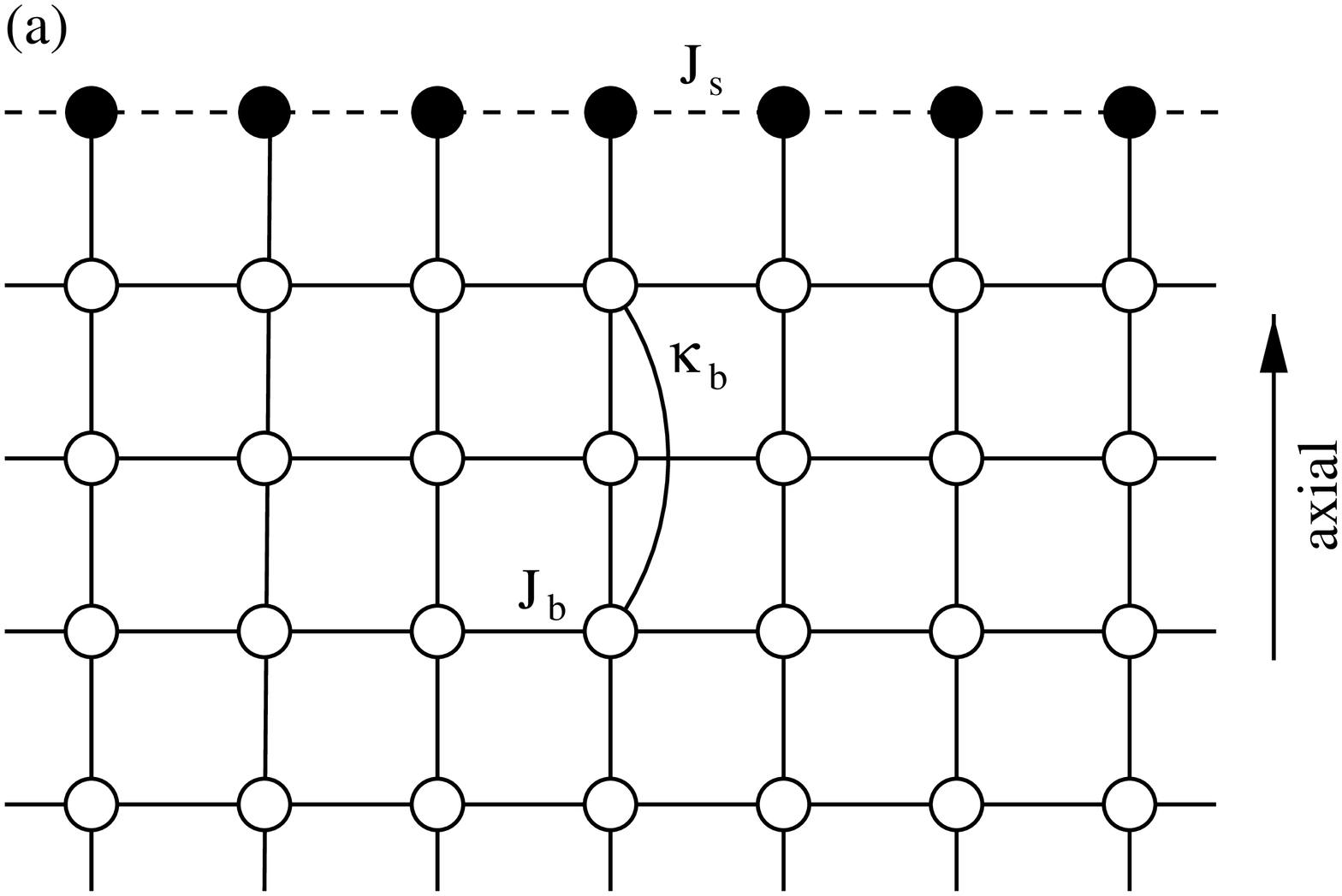,width=3.3in,angle=270}}
\vspace*{0.3cm}
\centerline{\psfig{figure=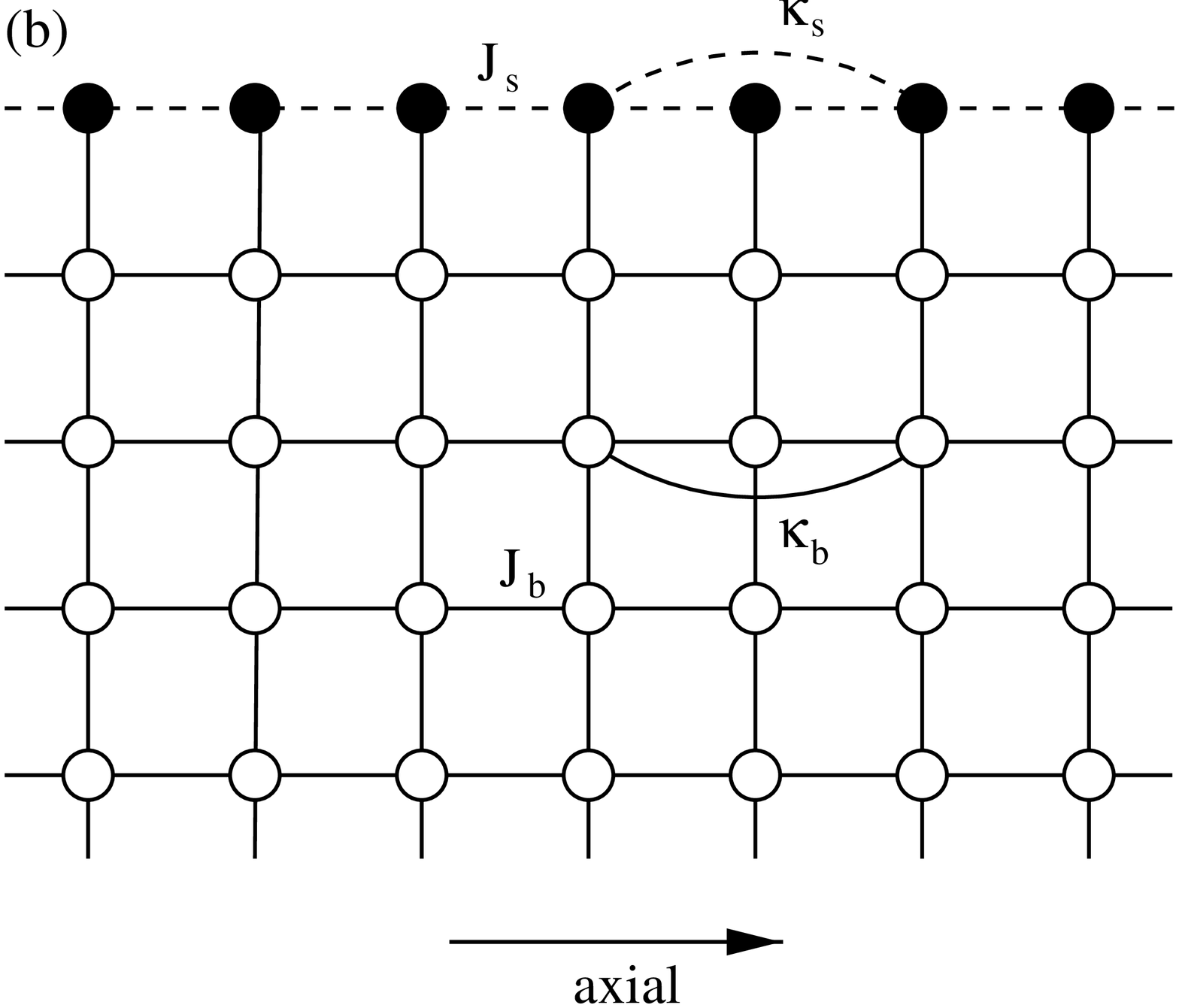,width=3.0in,angle=270}}
\caption{Cross sections of semi-infinite three-dimensional ANNNI models 
showing
the two different types of surfaces studied in the present work:
(a) surfaces perpendicular to the axis of competing interactions, (b)
surfaces parallel to this axis. $J_b$ and $J_s$ denote the nearest neighbor
bulk and surface couplings, respectively, whereas the axial next-nearest
neighbor interactions are labeled by the bulk, $\kappa_b$, and surface, $\kappa_s$,
competition parameters. Surface lattice sites are represented by filled points.
\label{fig1}} \end{figure}

\begin{figure}
\centerline{\psfig{figure=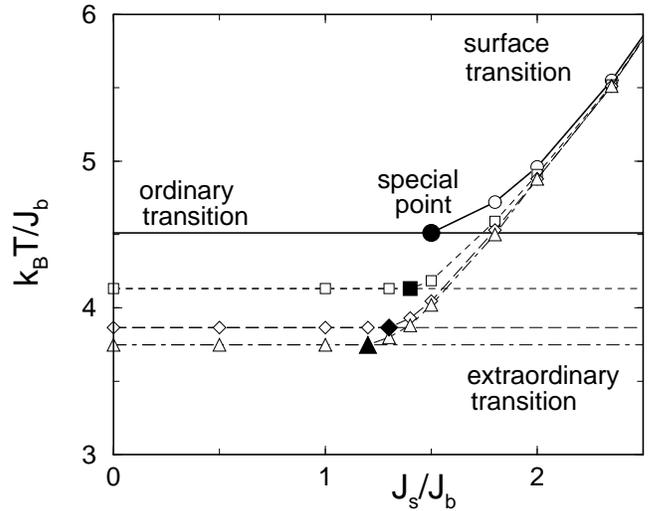,width=3.3in}}
\caption{Surface phase diagrams obtained for different values of the bulk competition
parameter $\kappa_b$: 0 (circles), 0.15 (squares), 0.24 (diamonds), and 0.27 (triangles).
Computed critical temperatures are shown as symbols, the filled symbols indicate the
locations of the different special transition points.
The shift in the bulk critical temperatures (horizontal lines) as a function of $\kappa_b$ is obvious.
Note that for the Lifshitz point value of $\kappa_b$ ($\kappa_b^L=0.27$) a special transition point
and a surface transition line are still encountered.
\label{fig2}} \end{figure}

\begin{figure}
\centerline{\psfig{figure=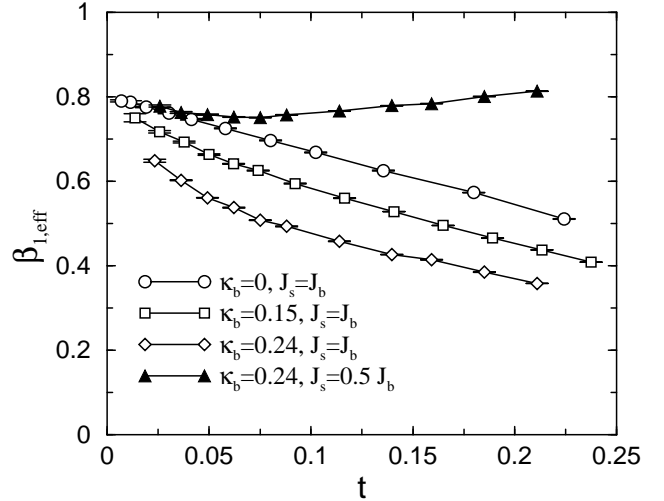,width=3.3in}}
\caption{Surface magnetization effective exponents as function of the reduced temperature $t$ for different
interaction sets as indicated in the legend. Note that in the limit $t \longrightarrow 0$
all curves extrapolate to the common
asymptotic value $\beta_1 \approx 0.80$, thus demonstrating that the ordinary transitions encountered 
for these interaction sets all belong to the same Ising surface universality class. 
In order to avoid finite-size dependences Monte Carlo systems with up to $120 \times 120 \times 40$ spins have
been simulated. Error bars result from averaging over at least ten different realizations.
\label{fig3}} \end{figure}

\begin{figure}
\centerline{\psfig{figure=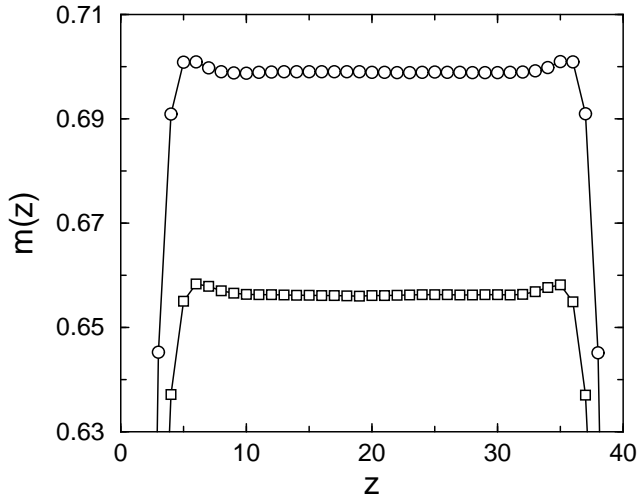,width=3.3in}}
\caption{Layer magnetizations obtained in the Monte Carlo simulation of an ANNNI film containing $120 \times 120
\times 40$ spins, with $\kappa_b=\kappa_b^L$ and $J_s = 0.75 J_b$. The two data sets correspond to two different 
temperatures: $k_BT/J_b=3.55$ (circles) and $k_BT/J_b=3.60$ (squares). Shown is the most interesting part where
a non-monotonic behaviour as function of the layer index is observed. The error bars are far smaller than the 
sizes of the symbols.
\label{fig4}} \end{figure} 

\begin{figure}
\centerline{\psfig{figure=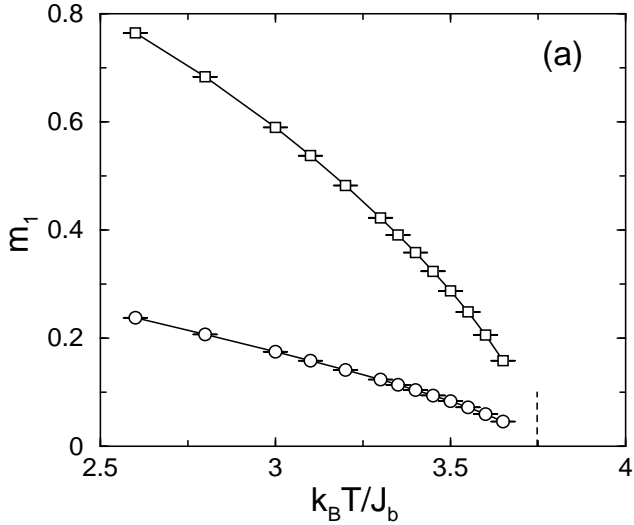,width=3.3in}}
\centerline{\psfig{figure=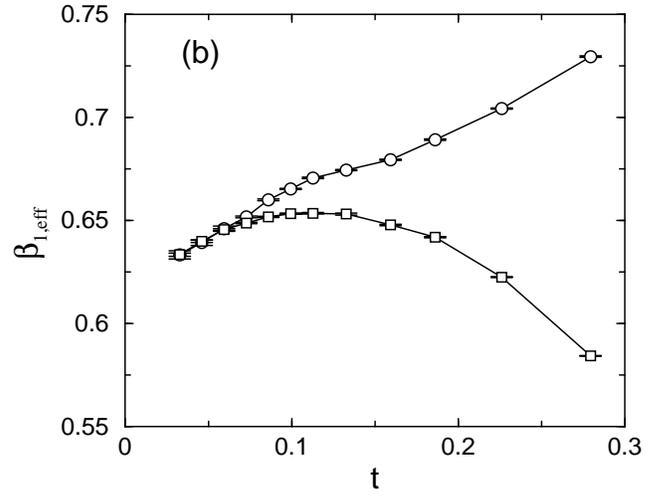,width=3.3in}}
\caption{(a) Surface magnetization versus temperature and (b) effective exponent $\beta_{1,eff}$ versus $t$ for $\kappa_b
= \kappa_b^L$. The data shown have been obtained for two values of the surface couplings: $J_s=0$ (circles) and $J_s=0.75 J_b$
(squares). A linear extrapolation of the effective exponents close to $t=0$ yields the estimate $\beta_1=0.618 \pm 0.005$
for the surface magnetization critical exponent at the Lifshitz point. In (a) the dashed line indicates the Lifshitz
point critical temperature $k_BT_c^L/J_b=3.7475$. Only data not affected by finite-size effects are shown.
Systems with up to $120 \times 120 \times 60$ spins have been simulated.
\label{fig5}} \end{figure} 

\begin{figure}
\centerline{\psfig{figure=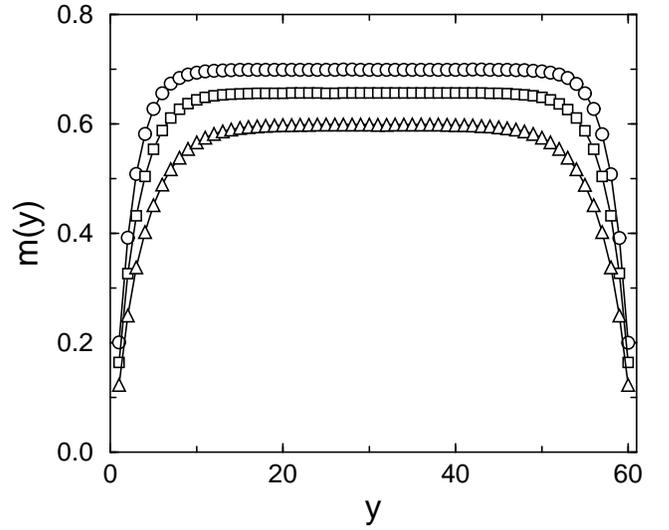,width=3.3in}}
\caption{Layer magnetizations computed for ANNNI samples with surfaces parallel to the
axial direction at different temperatures: $k_BT/J_b=3.55$ (circles), $k_BT/J_b=3.60$ (squares)
and $k_BT/J_b=3.65$ (triangles). Films with $60 \times 60 \times 60$ spins have been simulated,
with $\kappa_s=\kappa_b=\kappa_b^L$ and $J_s=J_b$. The layer magnetization increases
monotonically with the distance to the surface. Error bars are far smaller than the
sizes of the symbols.
\label{fig6}} \end{figure}

\begin{figure}
\centerline{\psfig{figure=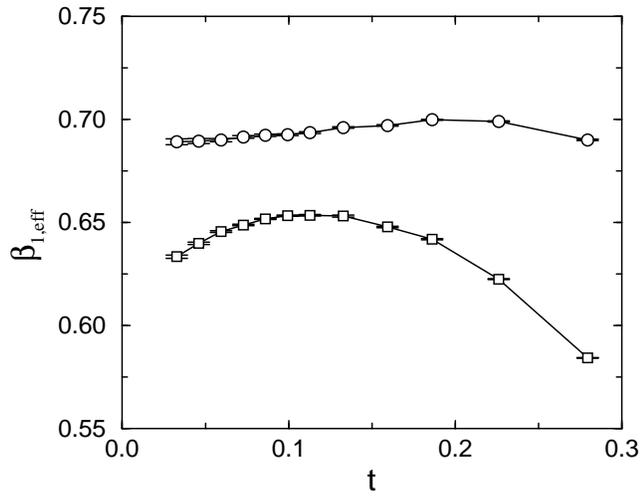,width=3.3in}}
\caption{Comparison of surface magnetization effective exponents obtained at the bulk Lifshitz point
from ANNNI samples with different surface orientations: surface perpendicular to the axial direction
(squares) or parallel to that axis (circles). In the former case $J_s=0.75 J_b$, whereas
in the latter case $\kappa_s=\kappa_b=\kappa_b^L$ and $J_s=J_b$. Extrapolating these data to
$t=0$ yields different values for the Lifshitz point surface magnetization critical exponent at the 
ordinary transition. In order to circumvent finite-size effects, different system sizes have been
simulated. The largest system considered contained $120 \times 120 \times 60$ spins (squares) or 
$80 \times 80 \times 80$ spins (circles), respectively.
\label{fig7}} \end{figure}

\end{document}